\documentclass[11pt,onecolumn]{IEEEtran}

\usepackage{amsmath,amsfonts, amssymb,color, mathrsfs, cite}
\usepackage{tikz, subcaption}
\usepackage[font={small,it}]{caption}
\usetikzlibrary{decorations.pathmorphing}
\usetikzlibrary{decorations.pathreplacing}
\usepackage[labelformat=simple]{subcaption}

\newtheorem{thm}{Theorem}

\newtheorem{corollary}{Corollary}
\newtheorem{lem}{Lemma}
\newtheorem{remark}{Remark}
\newtheorem{definition}{Definition}
\newtheorem{example}{Example}
\newtheorem{proposition}{Proposition}

\ifCLASSOPTIONcompsoc
  \usepackage[nocompress]{cite}
\else
  \usepackage{cite}
\fi
\begin{document}
%
%
%
%
%

\title{\LARGE \bf Robustness and Algebraic Connectivity of Random Interdependent Networks}
\author{Ebrahim Moradi Shahrivar,~Mohammad Pirani~and~Shreyas Sundaram
\IEEEcompsocitemizethanks{\IEEEcompsocthanksitem Ebrahim Moradi Shahrivar is with the Department of Electrical and Computer Engineering at the University of Waterloo. E-mail: {\tt emoradis@uwaterloo.ca}.
\IEEEcompsocthanksitem Mohammad Pirani is with the Department of Mechanical and Mechatronic Engineering at the University of Waterloo. E-mail: {\tt mpirani@uwaterloo.ca}. 
\IEEEcompsocthanksitem Shreyas Sundaram is with the School of Electrical and Computer Engineering at Purdue University.  E-mail: {\tt sundara2@purdue.edu}. Corresponding author.}

\thanks{This material is based upon work supported in part by the Natural Sciences and Engineering Research Council of Canada.}
}

\IEEEtitleabstractindextext{%
\begin{abstract}
We investigate certain structural properties of random interdependent networks.  We start by studying a property known as $r$-robustness, which is a strong indicator of the ability of a network to tolerate structural perturbations and dynamical attacks.  We show that random $k$-partite graphs exhibit a threshold for $r$-robustness, and that this threshold is the same as the one for the graph to have minimum degree $r$.  We then extend this characterization to random interdependent networks with arbitrary intra-layer topologies.  Finally, we characterize the algebraic connectivity of such networks, and provide an asymptotically tight rate of growth of this quantity for a certain range of inter-layer edge formation probabilities.  Our results arise from a characterization of the isoperimetric constant of random interdependent networks, and yield new insights into the structure and robustness properties of such networks.
\end{abstract}

}

\maketitle

\IEEEdisplaynontitleabstractindextext

%
\IEEEpeerreviewmaketitle


%
%
%
%

\section{Introduction}\label{sec:intro}
 There is an increasing realization that many large-scale networks are really  ``networks-of-networks,'' consisting of interdependencies between different subnetworks (e.g., coupled cyber and physical networks) \cite{P15, P17, MoradiShahrivar13, Radicchi, Hernandez,yazdani}. 
 Due to the prevalence of such networks, their robustness to intentional disruption or natural malfunctions has started to attract attention by a variety of researchers \cite{P16, yazdani,yagan,Marzieh13}.  
 In this paper, we contribute to the understanding of interdependent networks by studying the graph-theoretic properties of {\it $r$-robustness} and {\it algebraic connectivity} in such networks.  As we will describe further in the next section, $r$-robustness has strong connotations for the ability of networks to withstand structural and dynamical disruptions:  it guarantees that the network will remain connected even if up to $r-1$ nodes are removed from the neighborhood of {\it every} node in the network, and facilitates certain consensus dynamics that are resilient to adversarial nodes  \cite{Hogan,yagan2,dibaji2015consensus,hogan2, vaidya}.   The algebraic connectivity of a network is the second smallest eigenvalue of the Laplacian of that network, and plays a key role in the speed of certain diffusion dynamics \cite{Olfati1}.

We focus our analysis on a class of  {\it random interdependent networks} consisting of $k$ {\it layers} (or subnetworks), where each edge between nodes in different layers is present independently with certain probability $p$.  Our model is fairly general in that we make no assumption on the topologies within the layers, and captures Erdos-Renyi graphs and random $k$-partite graphs as special cases. We identify a bound $p_r$ for the probability of inter-layer edge formation $p$ such that for $p>p_r$, random interdependent networks with arbitrary intra-layer topologies are guaranteed to be $r$-robust asymptotically almost surely.  For the special case of $k$-partite random graphs, we show that this $p_r$ is tight (i.e., it forms a threshold for the property of $r$-robustness), and furthermore, is {\it also} the threshold for the minimum degree of the network to be $r$.  This is a potentially surprising result, given that $r$-robustness is a significantly stronger graph property than $r$-minimum-degree.   Recent work has shown that these properties also share thresholds in Erdos-Renyi random graphs \cite{hogan2} and random intersection graphs \cite{yagan2}, and our work in this paper adds random $k$-partite graphs to this list.

 Next, we show that when the inter-layer edge formation probability $p$ satisfies a certain condition, both the robustness parameter and the algebraic connectivity of the network grow as $\Theta(np)$ asymptotically almost surely (where $n$ is the number of nodes in each layer), {\it regardless} of the topologies within the layers.   
 Given the key role of algebraic connectivity in the speed of consensus dynamics on networks \cite{Olfati1}, our analysis demonstrates the importance of the edges that connect different communities in the network in terms of facilitating information spreading, in line with classical findings in the sociology literature \cite{Granovetter}.  Our result on algebraic connectivity of random interdependent networks is also applicable to the stochastic block model or planted partition model that has been widely studied in the machine learning literature  
 \cite{FK0, Abbe, Dasgupta, McSherry}. While we consider arbitrary intra-layer topologies, in the planted partition model it is assumed that the intra-layer edges are also placed randomly with a certain probability.   Furthermore, the lower bound that we obtain here for $\lambda_2(L)$ is tighter than the lower bounds obtained in the existing planted partition literature for the range of edge formation probabilities that we consider \cite{FK0, Dasgupta}. Both our robustness and algebraic connectivity bounds arise from a characterization that we provide of the isoperimetric constant of random interdependent networks.

\section{Graph Definitions and Background}
\label{sec:defs}
An undirected graph is denoted by $G = (V, E)$ where $V$ is the set of vertices (or nodes) and $E \subseteq V\times V$ is the set of edges. We denote the set $\mathcal{N}_i = \{v_j \in V~|~(v_i, v_j) \in E\}$ as the {\it neighbors} of  node $v_i \in V$ in graph $G$.  The {\it degree} of node $v_i$ is $d_i = |\mathcal{N}_i|$, and $d_{min}$ and $d_{max}$ are the minimum and maximum degrees of the nodes in the graph, respectively. A graph $G'= (V', E')$ is called a subgraph of $G=(V, E)$, denoted as $G' \subseteq G$, if $V' \subseteq V$ and $E' \subseteq E \cap \{ V' \times V' \}$. 
For an integer $k \in \mathbb{Z}_{\ge 2}$, a graph $G$ is $k$-partite if its vertex set can be partitioned into $k$ sets $V_1, V_2, \ldots, V_k$ such that there are no edges between nodes within any of those sets.  

The {\it edge-boundary} of a set of nodes $S \subset V$ is given by $\partial{S} = \{(v_i,v_j) \in E \mid v_i \in S, v_j \in V\setminus{S}\}$.  The {\it isoperimetric constant} of $G$ is defined as \cite{ChungSpectral}
 \begin{equation}
 i(G)\triangleq \min_{A \subset V, |A| \le \frac{n}{2}}\frac{|\partial A|}{|A|}.
 \label{eqn:iso}
 \end{equation}
By choosing $A$ as the vertex with the smallest degree we obtain $i(G) \le d_{min}$.
 
 The {\it adjacency matrix} for the graph is a matrix $A \in \{0,1\}^{n\times{n}}$ whose $(i,j)$ entry is $1$ if $(v_i,v_j) \in \mathcal{E}$, and zero otherwise.  The {\it Laplacian matrix} for the graph is given by $L = D - A$, where $D$ is the degree matrix with $D = \text{diag}(d_1, d_2, \ldots, d_n)$.  For an undirected graph $G$, the Laplacian $L$ is a symmetric matrix with real eigenvalues that can be ordered sequentially as $0=\lambda_1 (L)\leq \lambda_2 (L)\leq \cdots \leq \lambda_n (L)\leq 2d_{max}$.  The second smallest eigenvalue $\lambda_2(L)$ is termed the {\it algebraic connectivity} of the graph  and satisfies the bound \cite{ChungSpectral}
 \begin{equation}
\frac{i(G)^2}{2d_{max}} \le \lambda_2(L)  \le 2 i(G).
\label{eqn:lower_bound_lambda_2_iso}
\end{equation}

Finally, we will use the following consequence of the Cauchy-Schwartz inequality:
\begin{align}\label{cauchysw}
\sum_{i=1}^{k} s_i^2 \ge \frac{\big(\sum_{i=1}^{k} s_i\big)^2}{k},
\end{align}
where $s_i \in \mathbb{R}$ for $1 \le i \le k$.

\subsection{$\lowercase{r}$-Robustness of Networks}
\label{robustness}

Early work on the robustness of networks to structural and dynamical disruptions focused on the notion of {\it graph-connectivity}, defined as the smallest number of nodes that have to be removed to disconnect the network \cite{west01}. A network is said to be $r$-connected if it has connectivity at least $r$. 
In this paper, we will consider a stronger metric known as $r$-robustness, given by the following definition.

\begin{definition}[\cite{Hogan}]
Let $r \in \mathbb{N}$.  A subset $S$ of nodes in the graph $G=(V, E)$ is said to be {\it $r$-reachable} if there exists a node $v_i \in S$ such that $|\mathcal{N}_i\setminus S| \ge r$.  A graph $G=(V, E)$ is said to be {\it $r$-robust} if for every pair of nonempty, disjoint subsets of $V$, at least one of them is $r$-reachable.
\end{definition} 

Simply put, an $r$-reachable set contains a node that has $r$ neighbors outside that set, and an $r$-robust graph has the property that no matter how one chooses two disjoint nonempty sets, at least one of those sets is $r$-reachable.  This notion carries the following important implications:  

\begin{itemize} 
\item If network $G$ is $r$-robust, then it is at least $r$-connected and has minimum degree at least $r$ \cite{Hogan}. 
\item An $r$-robust network remains connected even after removing up to $r-1$ nodes from the neighborhood of {\it every} remaining node \cite{hogan2}.
\item Consider the following class of consensus dynamics where each node starts with a scalar real value.  At each iteration, it discards the highest $F$ and lowest $F$ values in its neighborhood (for some $F\in\mathbb{N}$), and updates its value as a weighted average of the remaining values.  It was shown in \cite{Hogan,dibaji2015consensus} that if the network is $(2F+1)$-robust, all nodes that follow these dynamics will reach consensus  even if there are up to $F$ arbitrarily behaving malicious nodes in the neighborhood of {\it every} normal node. 
\end{itemize}

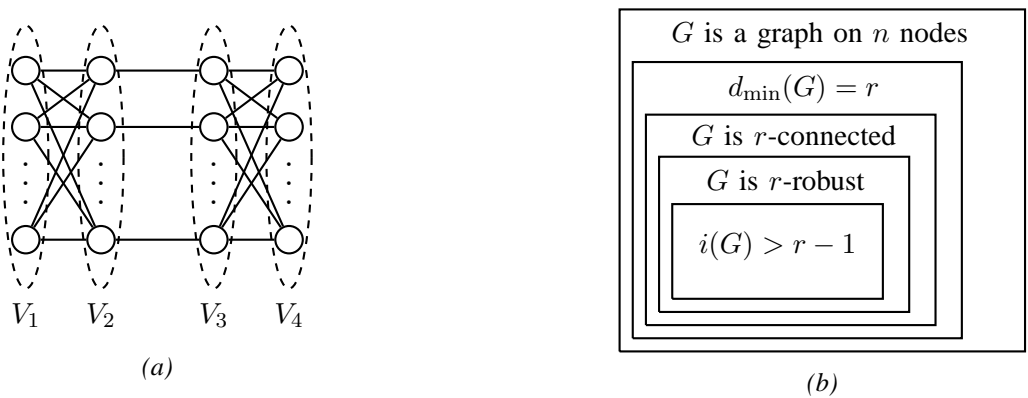
\begin{figure}[b]
\begin{minipage}{0.48\textwidth}
\begin{center}
\begin{tikzpicture}[scale=.50, auto, node distance=1cm, thick,
   node_style/.style={circle,draw=black,fill=white!20!,font=\sffamily\Large\bfseries},
   edge_style/.style={draw=black, thick}]

  \node[node_style] (n1) at (1,3.5)  {};
  \node (n10) at (1,4.5)  {{\Large $.$}};
  \node (n10) at (1,5)  {{\Large $.$}};
  \node (n10) at (1,5.5)  {{\Large $.$}};
  \node[node_style](n2) at (1,6.5)  {};
  \node[node_style] (n3) at (1,8)  {};

  \node[node_style] (n4) at (3,3.5)  {};
  \node (n10) at (3,4.5)  {{\Large $.$}};
  \node (n10) at (3,5)  {{\Large $.$}};
  \node (n10) at (3,5.5)  {{\Large $.$}};
  \node[node_style](n5) at (3,6.5)  {};
  \node[node_style] (n6) at (3,8)  {};

  \node[node_style] (n7) at (6,3.5)  {};
  \node (n10) at (6,4.5)  {{\Large $.$}};
  \node (n10) at (6,5)  {{\Large $.$}};
  \node (n10) at (6,5.5)  {{\Large $.$}};
  \node[node_style](n8) at (6,6.5)  {};
  \node[node_style] (n9) at (6,8)  {};

  \node[node_style] (n11) at (8,3.5)  {};
  \node (n10) at (8,4.5)  {{\Large $.$}};
  \node (n10) at (8,5)  {{\Large $.$}};
  \node (n10) at (8,5.5)  {{\Large $.$}};
  \node[node_style](n12) at (8,6.5)  {};
  \node[node_style] (n13) at (8,8)  {};

\draw (n1) --(n4);
\draw (n1) --(n5);
\draw (n1) --(n6);

\draw (n2) --(n4);
\draw (n2) --(n5);
\draw (n2) --(n6);

\draw (n3) --(n4);
\draw (n3) --(n5);
\draw (n3) --(n6);

\draw (n7) --(n11);
\draw (n7) --(n12);
\draw (n7) --(n13);

\draw (n8) --(n11);
\draw (n8) --(n12);
\draw (n8) --(n13);

\draw (n9) --(n11);
\draw (n9) --(n12);
\draw (n9) --(n13);

\draw (n4) --(n7);
\draw (n5) --(n8);
\draw (n6) --(n9);

 \draw (1,5.7) ellipse (0.6cm and 3.5cm) [dashed];
 \draw (3,5.7) ellipse (0.6cm and 3.5cm) [dashed];
 \draw (6,5.7) ellipse (0.6cm and 3.5cm) [dashed];
 \draw (8,5.7) ellipse (0.6cm and 3.5cm) [dashed];

  \node at (1,1.5) {$V_1$};
   \node at (3,1.5) {$V_2$};
  \node at (6,1.5) {$V_3$};
  \node at (8,1.5) {$V_4$};



\end{tikzpicture}    
    \subcaption{}
    \label{fig1}
\end{center}
\end{minipage}
\begin{minipage}{0.48\textwidth}
\begin{center}
\begin{tikzpicture}[scale=.35, auto, node distance=5cm, thick,
   node_style/.style={circle,draw=black,fill=white!20!,font=\sffamily\Large\bfseries},
   edge_style/.style={draw=black, thick}]

\draw (-7,-3.5)--(1,-3.5)--(1,0.1)--(-7,0.1)--(-7,-3.5);
\node at (-3,-1.5) {$i(G) >r-1$};
\draw (-7.5,-4)--(2,-4)--(2,1.9)--(-7.5,1.9)--(-7.5,-4);
\node at (-2.7,1) {$G$ is $r$-robust};
\draw (-8,-4.5)--(3,-4.5)--(3,3.5)--(-8,3.5)--(-8,-4.5);
\node at (-2.4,2.7) {$G$ is $r$-connected};
\draw (-8.5,-5)--(4,-5)--(4,5.5)--(-8.5,5.5)--(-8.5,-5);
\node at (-2,4.5) {$d_{\min}(G)=r$};
\draw (-9,-5.5)--(6.4,-5.5)--(6.4,7.5)--(-9,7.5)--(-9,-5.5);
\node at (-1.4,6.5) {$G$ is a graph on $n$ nodes};



\end{tikzpicture}    
    \subcaption{}
    \label{fig2}
\end{center}
\end{minipage}
\caption{(a) Graph $G=(V, E)$ with $V=V_1\cup V_2\cup V_3\cup V_4$ and $|V_i|=\frac{n}{4},~ 1 \le i \le 4$. All of the nodes in $V_1$ ($V_3$) are connected to all of the nodes in $V_2$ ($V_4$). Furthermore, there is a one to one connection between nodes in $V_2$ and nodes in $V_3$.~(b) Relationships between different notions of robustness.}
\end{figure}

Thus, $r$-robustness is a stronger property than $r$-minimum-degree and $r$-connectivity.  Indeed, the gap between the robustness and connectivity (and minimum degree) parameters can be arbitrarily large, as illustrated by the bipartite graph shown in Fig.~\ref{fig1}.  That graph has minimum degree $n/4$ and connectivity $n/4$.  However, if we consider the disjoint subsets $V_1 \cup V_2$ and $V_3 \cup V_4$, neither one of those sets contains a node that has more than $1$ neighbor outside its own set.  Thus, this graph is only $1$-robust.   

The following result shows that the isoperimetric constant $i(G)$ defined in \eqref{eqn:iso} provides a lower bound on the robustness parameter.

\begin{lem}\label{rob-iso}
Let $r$ be a positive integer. If $i(G)> r-1$, then the graph is at least $r$-robust.
\end{lem}

\begin{IEEEproof}
If $i(G) >r-1$, then every set of nodes $S \subset V$ of size up to $\frac{n}{2}$ has at least $(r-1)|S|+1$ edges leaving that set (by the definition of $i(G)$).  By the pigeonhole principle, at least one node in $S$ has at least $r$ neighbors outside $S$.  Now consider any two disjoint non-empty sets $S_1$ and $S_2$.  At least one of these sets has size at most $\frac{n}{2}$, and thus is $r$-reachable.  Therefore, the graph is $r$-robust.
\end{IEEEproof}
As an example of Lemma \ref{rob-iso}, consider the graph shown in Fig.~\ref{fig1}. Graph $G$ has isoperimetric constant of at most $0.5$ (since the edge boundary of $V_1 \cup V_2$ has size $n/4$), but is $1$-robust.  It is worth mentioning that it is possible to construct a graph such that difference between the robustness parameter and isoperimetric constant is arbitrarily large. For instance, given any arbitrary $t \in \mathbb{N}$ and $n$ sufficiently large, assume that the interconnection topology between partitions $V_2$ and $V_3$ in graph $G$ is $t$-regular (i.e., each node in $V_2$ ($V_3$) is connected to exactly $t$ nodes in $V_3$ ($V_2$)) and the rest of the graph is the same as the structure shown in Fig. \ref{fig1}. Then the isoperimetric constant of the graph $G$ is at most $t/2$ by considering the set $V_1 \cup V_2$. However, one can show that $G$ is $t$-robust. 

We have summarized the relationships  between these different graph-theoretic measures of robustness in Fig.~\ref{fig2}.  To give additional context to the properties highlighted above, we note the following.  Consider an arbitrary partition of the nodes of a graph into two sets.  An $r$-connected graph guarantees that  the nodes in one of the sets {\it collectively} have $r$ neighbors outside that set.  An $r$-robust graph guarantees that there is a node in one of the sets that {\it by itself} has $r$ neighbors outside that set.  A graph with $i(G) \ge r$ guarantees that {\it each node} in one of the sets has $r$ neighbors outside that set {\it on average}.

In the rest of the paper, we will study random interdependent networks and show how these various properties are related in such networks.  
\section{Random Interdependent Networks}
\label{sec:random_interdep_nets}
We start by formally defining the notion of (random) {\it interdependent networks} that we consider in this paper.

\begin{definition}
An interdependent network $G$  is denoted by a tuple $G=(G_1, G_2, \allowbreak \ldots, G_k, G_p)$ where $G_i=(V_i, E_i)$ for $i=1,2,\ldots, k$ are called the {\it layers} of the network $G$, and $G_p=(V_1\cup V_2 \cup \ldots \cup V_k, E_p)$ is a $k$-partite network with $E_p \subseteq \cup_{i \neq j} V_i \times V_j$ specifying the interconnection (or inter-layer) topology. 
\end{definition}

For the rest of this paper, we assume that $|V_1|=|V_2|=\cdots=|V_k|=n$ and the number of layers $k$ is at least 2.  

\begin{definition}
Define the probability space $(\Omega_n, \mathcal{F}_n, \mathbb{P}_n)$, where the sample space $\Omega_n$ consists of all possible interdependent networks $(G_1, G_2, \ldots, G_k, G_p)$ 
and the index $n \in \mathbb{N}$ denotes the number of nodes in each layer. The $\sigma$-algebra $\mathcal{F}_n$ is the power set of $\Omega_n$ and the probability measure $\mathbb{P}_n$ associates a probability $\mathbb{P}(G_1, G_2, \ldots, G_k, G_p)$ to each network $G=(G_1, G_2, \ldots, G_k, G_p)$. A random interdependent network is a network $G=(G_1, G_2, \ldots, G_k, G_p)$ drawn from $\Omega_n$ according to the given probability distribution. 
\end{definition}

Note that deterministic structures for the layers or interconnections can be obtained as a special case of the above definition where $\mathbb{P}(G_1, G_2, \ldots, G_k, G_p)$ is 0 for interdependent networks not containing those specific structures; for instance, a {\it random $k$-partite graph} is obtained by allocating a probability of $0$ to interdependent networks where  any of the $G_i$ for $1 \leq i \leq k$ is nonempty.  
Through an abuse of notation, we will refer to random $k$-partite graphs simply by $G_p$ in this paper.    Similarly, Erdos-Renyi random graphs  on $kn$ nodes are  obtained as a special case of the above definition by  choosing the edges in $G_1, G_2, \ldots, G_k$ and $G_p$ independently with a common probability $p$.

In this paper, we will focus on the case where $G_p$ is independent of $G_i$ for $1 \leq i \leq k$. Specifically, we assume that each possible edge of the $k$-partite network $G_p$ is present independently with probability $p$ (which can be a function of $n$).
We refer to this as a {\it random interdependent network with Bernoulli interconnections}.   We will be characterizing certain properties of such networks as $n$ gets large, captured by the following definition.

\begin{definition}
For a random interdependent network, we say that a property $\mathcal{P}$ holds {\it asymptotically almost surely} (a.a.s.) if the probability measure of the set of graphs with property $\mathcal{P}$  (over the probability space $(\Omega_n, \mathcal{F}_n, \mathbb{P}_n)$) tends to 1 as $n \to \infty$.
\end{definition}

\section{Robustness of Random interdependent networks}
We will first consider random $k$-partite networks (a special case of random interdependent networks, as explained in the previous section), and show that they exhibit phase transitions at certain thresholds for the probability $p$, defined as follows.

\begin{definition}
Consider a function $t(n)=\frac{g(n)}{n}$ with $g(n) \to \infty$ as $n \to \infty$, and a function $x=o(g(n))$ which satisfies $x \to \infty$ as $n \to \infty$. Then $t(n)$ is said to be a (sharp) threshold function for a graph property $\mathcal{P}$ if 
\begin{enumerate}
 \item property $\mathcal{P}$ a.a.s. holds when $p(n)=\frac{g(n)+x}{n}$, and
\item property $\mathcal{P}$ a.a.s. does not hold for $p(n)=\frac{g(n)-x}{n}$.
\end{enumerate}
\end{definition}

Loosely speaking, if the probability of adding an edge between layers is ``larger'' than the threshold $t(n)$, then property $\mathcal{P}$ holds; if $p$ is ``smaller'' than the threshold $t(n)$, then $\mathcal{P}$ does not hold.  The following theorem is one of our main results and characterizes the threshold for $r$-robustness (and $r$-minimum-degree) of random $k$-partite networks.

\begin{thm}\label{robust}
For any positive integers $r$ and $k \ge 2$,
$$
t(n)=\frac{\ln n +(r-1) \ln \ln n}{(k-1)n}
$$
is a threshold for $r$-robustness of random $k$-partite graphs with Bernoulli interconnections. It is also a threshold for the $k$-partite network to have a minimum degree $r$.
\end{thm}

In order to prove this theorem, we show the following stronger result (see  Fig.~\ref{fig2}) that $i(G) > r-1$ when the probability of edge formation is above the given threshold. 

\begin{lem}\label{lem:iso}
Consider a random $k$-partite graph $G_p=(V_1\cup V_2 \cup \cdots \cup V_{k}, E_p)$ with node sets $V_i=\{(i-1)n+1,(i-1)n+2,\ldots,in\}$ for $1 \leq i \leq k$. Assume that the edge formation probability $p=p(n)$ is such that $p(n)=\frac{\ln{n}+(r-1) \ln{\ln{n}}+x}{(k-1)n}$, where $r \in  \mathbb{N}$ is a constant such that $r \ge 1$ and $x=x(n)$ is some function satisfying $x=o(\ln{\ln{n}})$ and $x \to \infty$ as $n\to \infty$. Define $S_{r}$ as the property that for any set of nodes with size of at most $\lfloor kn/2 \rfloor$, we have $|\partial S| > (r-1) |S|$. Then $G_p$ has property $S_{r}$ a.a.s. In other words, we have $i(G_p) > r-1$ a.a.s.
\end{lem}
The proof of the above lemma is provided in Appendix \ref{appendixA}.
The idea behind the proof is to first use a union-bound to upper bound the probability that for some set $S$ of size $\lfloor kn/2 \rfloor$ or less, we have $|\partial S| \le |S|(r-1)$.  Via the use of algebraic manipulations and upper bounds, we then show that this probability goes to zero when the inter-edge formation probability $p(n)$ satisfies the conditions given in the lemma. 

Using the above lemma, we can prove Theorem~\ref{robust}.

\begin{IEEEproof}[Proof of Theorem \ref{robust}]
Consider  a random $k$-partite graph $G_p$ of the form described in the theorem with edge formation probability  $p(n)=\frac{\ln{n}+(r-1) \ln{\ln{n}}+x}{(k-1)n}$, where $r \in  \mathbb{N}$ is a constant and $x=x(n)$ is some function satisfying $x=o(\ln{\ln{n}})$ and $x \to \infty$ as $n\to \infty$. By Lemma \ref{lem:iso}, we know that $i(G) >r-1$. Therefore, by Lemma \ref{rob-iso}, the random $k$-partite graph $G_p$ is at least $r$-robust a.a.s. This also implies that it has minimum degree at least $r$ a.a.s. (by the relationships shown in Fig.~\ref{fig2}).

Next we have to show that for $p(n)=\frac{\ln{n}+(r-1) \ln{\ln{n}}-x}{(k-1)n}$, where $x=x(n)$ is some function satisfying $x=o(\ln{\ln{n}})$ and $x \to \infty$ as $n\to \infty$, a random $k$-partite graph is asymptotically almost surely not $r$-robust. Let $G_p=(V_1\cup V_2 \cup \cdots \cup V_{k}, E_p)$ be a random $k$-partite graph.  Consider the vertex set $V_1 = \{v_1, \ldots, v_n\}$, and define the random variable $X=X_1+X_2+\cdots+X_n$ where $X_i=1$ if the degree of node $v_i$ is less than $r$ and zero otherwise. The goal is to show that if $p(n)=\frac{\ln{n}+(r-1) \ln{\ln{n}}-x}{(k-1)n}$, 
then $\mathbf{Pr}(X=0) \to 0$ asymptotically almost surely. This means that for the specified $p(n)$, there exists a node in $V_1$ with degree less than $r$ with high probability.  Since any $r$-robust graph must have minimum degree of at least $r$, we will have the required result.

The random variable $X$ is zero if and only if  $X_i = 0$ for $1 \leq i \leq n$. The random variables $X_i$ and $X_j$  are identically distributed and  independent when $i \neq j$ and thus we have 
\begin{align}
\mathbf{Pr}(X=0)=\mathbf{Pr}(X_1=0)^n =(1-\mathbf{Pr}(X_1=1))^n \leq e^{-n\mathbf{Pr}(X_1=1)},
\label{eqn:PX0_bound}
\end{align}
where the last inequality is due to the fact that $1-p \leq e^{-p}$ for $p\ge 0$. 
Now, note that
\begin{align}
n\mathbf{Pr}(X_1=1)&=n\sum_{i=0}^{r-1} {n(k-1) \choose i}p^i (1-p)^{n(k-1)-i} \nonumber \\
&\geq n {n(k-1) \choose r-1} p^{r-1} (1-p)^{n(k-1)-r+1} \nonumber \\
&\geq n {n(k-1) \choose r-1} p^{r-1} (1-p)^{n(k-1)}, \label{eqn:boundnP}
\end{align}
where the last inequality is obtained by using the fact that  $0 < (1-p)^{r-1} \le 1$ for $r \ge 1$.  Using the fact that ${n(k-1) \choose r-1} = \Omega\left(n^{r-1}\right)$ for constant $r$ and $k$, and $(1-p)^{n(k-1)}=e^{n(k-1)\ln(1-p)} = \Omega(e^{-n(k-1)p})$ when $np^2 \rightarrow 0$ (which is satisfied for the function $p$ that we are considering above), the inequality \eqref{eqn:boundnP} becomes
$$
n\mathbf{Pr}(X_1=1) =\Omega\left(n^rp^{r-1}e^{-n(k-1)p}\right).
$$
Substituting $p =\frac{\ln{n}+(r-1) \ln{\ln{n}}-x}{(k-1)n}$ and simplifying, we obtain
\begin{align*}
n\mathbf{Pr}(X_1=1) =\Omega\left( \frac{(\ln{n}+(r-1) \ln{\ln{n}}-x)^{r-1}}{ (\ln n)^{r-1}} e^x\right)= \Omega(e^x).
\end{align*}

Thus we must have that $\lim_{n\to \infty} n\mathbf{Pr}(X_1=1)=\infty$, which proves that $\mathbf{Pr}(X=0) \to 0$ as $n \to \infty$ (from \eqref{eqn:PX0_bound}). Therefore, there exists a node with degree less than $r$ and thus the random $k$-partite graph $G_p$ is not $r$-robust for $p(n)=\frac{\ln{n}+(r-1) \ln{\ln{n}}-x}{(k-1)n}$. 
\end{IEEEproof}

As described in the introduction, the above result indicates that the properties of $r$-robustness and $r$-minimum-degree (and correspondingly, $r$-connectivity) all share the same threshold function in random $k$-partite graphs, despite the fact that $r$-robustness is a significantly stronger property than the other two properties.  In particular,  this indicates that above the given threshold, random $k$-partite networks possess stronger robustness properties than simply being $r$-connected:  they can withstand the removal of a large number of nodes (up to $r-1$ from every neighborhood), and facilitate purely local consensus dynamics that are resilient to a large number of malicious nodes (up to $\lfloor\frac{r-1}{2}\rfloor$ in the neighborhood of every normal node).

\subsection{General Random Interdependent Networks}
With the sharp threshold given by Theorem~\ref{robust} for random $k$-partite graphs in hand, we now consider general random interdependent networks with arbitrary topologies within the layers.  Note that any general random interdependent network can be obtained by first drawing a random $k$-partite graph, and then adding additional edges to fill out the layers.  Using the fact that $r$-robustness is a {\it monotonic graph property} (i.e., adding edges to an $r$-robust graph does not decrease the robustness parameter),  we obtain the following result.

\begin{corollary}
Consider a random interdependent graph  $G=(G_1, G_2, \ldots, G_{k}, \allowbreak G_p)$ with Bernoulli interconnections. Assume that  the inter-layer edge formation probability satisfies $p(n) \ge \frac{\ln{n}+(r-1) \ln{\ln{n}}+x}{(k-1)n}$, $r \in  \mathbb{Z}_{\ge 1}$ and $x=x(n)$ is some function satisfying $x=o(\ln{\ln{n}})$ and $x \to \infty$ as $n\to \infty$.  Then $G$ is $r$-robust.
\end{corollary}

The above result shows that if the inter-layer edge formation probability between all layers of the $k$-layer graph $G$ is higher than the threshold for $r$-robustness of a $k$-partite network, then $G$ is a $r$-robust network, {\it regardless} of the probability distribution over the topologies within the layers.  

\subsection{Unbounded Robustness in Random Interdependent Networks}
The previous results (Theorem~\ref{robust} in particular) established inter-layer edge formation probabilities that cause random interdependent networks to be $r$-robust, and demonstrated that the properties of $r$-minimum-degree, $r$-conne -ctivity and $r$-robustness share the same probability threshold in random $k$-partite  graphs (see Fig.~\ref{fig2}). 
Here, we will investigate a coarser rate of growth for the inter-layer edge formation $p$, and show that for such probability functions, the isoperimetric constant and the robustness parameter have the same asymptotic rate of growth. This will also play a role in the next section, where we investigate the algebraic connectivity of random interdependent networks.  We start with the following lemma.

\begin{lem}\label{lem:iso_bounds}
Consider a random $k$-partite graph $G_p=(V_1\cup V_2 \cup \cdots \cup V_{k}, E_p)$ with node sets $V_i=\{(i-1)n+1,(i-1)n+2,\ldots,in\}$ for $1 \leq i \leq k$.
Assume that the edge formation probability $p$ satisfies $\lim \sup_{n \to \infty} \frac{\ln n}{(k-1)n p} <1$. Fix any $\epsilon \in (0, \frac{1}{2}]$. There exists a constant $\alpha$ (that depends on $p$) such that the minimum degree $d_{min}$, maximum degree $d_{max}$ and isoperimetric constant $i(G_p)$ a.a.s. satisfy
\begin{equation}\label{prop_e_1}
\alpha n p \leq i(G_p) \leq d_{min} \leq d_{max}\leq  n(k-1)p \left( 1+\sqrt{3} \left(\frac{\ln n}{(k-1)n p}\right)^{\frac{1}{2}-\epsilon}\right).
\end{equation}
\end{lem}

The proof of the above lemma is given in Appendix \ref{appendixB}. The above result leads to the following theorem.

\begin{thm}\label{infrobust}
Consider a random $k$-partite graph  $G$ with Bernoulli interconnections. Assume that  the inter-layer edge formation probability $p = p(n)$ satisfies 
$$
\lim \sup_{n \to \infty} \frac{\ln n}{(k-1)n p} <1.
$$  
Then $i(G) = \Theta(np)$ a.a.s., and furthermore, $G$ is $\Theta(np)$-robust a.a.s.
\end{thm}

\begin{IEEEproof}
For inter-layer edge formation probabilities satisfying the given condition, we have $i(G) = \Theta(np)$ a.a.s. from Lemma~\ref{lem:iso_bounds}.  From Lemma~\ref{rob-iso}, we have that the robustness parameter is $\Omega(np)$ a.a.s.  Furthermore, since the robustness parameter is always less than the minimum degree of the graph, the robustness parameter is $O(np)$ a.a.s. from Lemma~\ref{lem:iso_bounds}, which concludes the proof.
\end{IEEEproof}

Once again, since adding edges to a network cannot decrease the isoperimetric constant or the robustness parameter, the above result immediately implies that for random interdependent networks with inter-layer edge formation probability satisfying $\lim \sup_{n \to \infty} \frac{\ln n}{(k-1)n p} <1$, we have $i(G) = \Omega(np)$ and that the robustness is $\Omega(np)$ a.a.s.  This condition on the inter-layer edge formation probability has further implications for the structure of random interdependent networks. In next section, we use Lemma~\ref{lem:iso_bounds} to show that the algebraic connectivity of random interdependent networks  scales as $\Theta(np)$ a.a.s. for all $p$ that satisfy the condition, again  regardless of the probability distributions over the topologies of the layers.
\section{Algebraic Connectivity of Random Interdependent Networks}
\label{algebraic}
The algebraic connectivity of  interdependent networks has started to receive attention in recent years.  The authors of \cite{Radicchi} analyzed the algebraic connectivity of deterministic  interconnected networks with one-to-one weighted symmetric inter-layer connections. The recent paper by Hernandez et al. studied the algebraic connectivity of a mean field model of interdependent networks where each layer has an identical structure, and the interconnections are all-to-all with appropriately chosen weights \cite{Hernandez}. Spectral properties of random interdependent networks (under the moniker of {\it planted partition models}) have also been studied in research areas such as algorithms and machine learning \cite{Abbe, McSherry, Newman, Kwok, FK0}.  
Here, we leverage our results from the previous section to provide a bound on the algebraic connectivity for random interdependent networks that, to the best of our knowledge, is the tightest known bound for the range of inter-layer edge formation probabilities that we consider. 

\begin{thm}
Consider a random interdependent graph  $G=(G_1, G_2, \ldots, G_{k}, G_p)$ with Bernoulli interconnections and assume that the probability  of inter-layer edge formation $p$ satisfies $\lim \sup_{n \to \infty} \frac{\ln n}{(k-1)n p} <1$. Then $\lambda_2(G)=\Theta(np)$ a.a.s.
\label{thm:alg_conn_random_interdep_graph}
\end{thm}

\begin{IEEEproof}
In order to prove this theorem, we need to show that there exist constants $\gamma, \beta >0$ such that $ \gamma np \le \lambda_2(G) \le \beta np$ a.a.s. We start with proving the existence of constant $\beta$. 

Consider the set of nodes $V_1$ in the first layer. The number of edges between $V_1$ and all other $V_j$, $2 \leq j \leq k$ is a binomial random variable $B(n^2(k-1), p)$ and thus $\mathbb{E} [|\partial V_1|]=n^2 (k-1) p$. By using the Chernoff bound \cite{Mitzenmacher} for the random variable $|\partial V_1|$, we have (for $0 < \delta <1$)  
\begin{align} \label{eqn:chernoff_eb}
\mathbf{Pr}(|\partial V_1| \geq (1+ \delta)\mathbb{E}[|\partial V_1|])\leq e^{\frac{-\mathbb{E}(|\partial V_1|)\delta^2}{3}}.
\end{align}
Choosing $\delta=\frac{\sqrt{3}}{\sqrt{\ln n}}$, the upper bound in the expression above becomes 
$\exp{\left(-\frac{n^2 (k-1)p}{\ln n}\right)}$. Since $\ln n < n (k-1) p$ for $n$ sufficiently large and for $p$ satisfying the condition in the proposition, the right hand side of inequality \eqref{eqn:chernoff_eb} goes to zero as $n \to \infty$. Thus $|\partial V_1| \leq (1+ o(1))\mathbb{E}[|\partial V_1|]$ a.a.s.  Therefore
\begin{align*}
i(G)=\operatorname*{min}_{\substack{|A| \leq \frac{nk}{2},\\
 A \subseteq V_1\cup V_2 \cup \cdots \cup V_{k}}} \frac{|\partial A|}{|A|} \leq \frac{|\partial V_1|}{|V_1|}\leq \frac{(1+o(1))n^2 (k-1) p}{n},
\end{align*}
a.a.s. Using \eqref{eqn:lower_bound_lambda_2_iso}, we have the required result.

Next, we prove the lower bound on $\lambda_2(G)$. Consider the $k$-partite subgraph of network $G$ which is denoted by $G_p$. By Lemma~\ref{lem:iso_bounds} and the inequality \eqref{eqn:lower_bound_lambda_2_iso}, we know that $\lambda_2(G_p) \geq \gamma n p$ for some constant $\gamma$ asymptotically almost surely.  Since adding edges to a graph does not  decrease the algebraic connectivity of that graph \cite{Brouwer}, we have $\lambda_2(G) \geq \lambda_2(G_p) \geq \gamma n p$ asymptotically almost surely.
\end{IEEEproof}

Theorem \ref{thm:alg_conn_random_interdep_graph} demonstrates the importance of inter-layer edges on the algebraic connectivity of the overall network when $\lim \sup_{n \to \infty} \frac{\ln n}{(k-1) n p} <1$.   This requirement on the growth rate of $p$ cannot be reduced if one wishes to stay agnostic about the probability distributions over the topologies of the layers.  Indeed, in the proof of the Theorem \ref{robust}, we showed that if $\lim \sup_{n \to \infty} \frac{\ln n}{(k-1)n p} > 1$, a random $k$-partite graph will have at least one isolated node asymptotically almost surely and thus has algebraic connectivity equal to zero asymptotically almost surely.   In this case the quantity $\frac{\ln{n}}{(k-1)n}$ forms a {\it coarse threshold} for the algebraic connectivity being $0$, or growing as $\Theta(np)$.  On the other hand, if one had further information about the probability distributions over the layers, one could potentially relax the condition on $p$ required in the above results.  For instance, as mentioned in Section~\ref{sec:random_interdep_nets}, when each of the $k$-layers is an Erdos-Renyi graph formed with probability $p$, then the entire interdependent network is an Erdos-Renyi graph on $kn$ nodes; in this case, the algebraic connectivity is $\Omega(np)$ asymptotically almost surely as long as $\lim \sup_{n \to \infty} \frac{\ln{n}}{knp} < 1$ \cite{TAC}.  This constraint on $p$ differs by a factor of $\frac{k}{k-1}$ from the expression in Theorem~\ref{thm:alg_conn_random_interdep_graph}.

\section{Summary and Future Work}
\label{sec:conc}
We studied the properties of $r$-robustness and algebraic connectivity in random interdependent networks.  We started by analyzing random $k$-partite networks, and showed that $r$-robustness and $r$-minimum-degree (and $r$-connectivity) all share the same threshold function, despite the fact that $r$-robustness is a much stronger property than the others.  This robustness carries over to random interdependent networks with arbitrary intra-layer topologies, and yields new insights into the structure of such networks (namely that they can tolerate the loss of a large number of nodes, and are resilient to misbehavior in certain dynamics).  We also provided tight asymptotic rates of growth on the algebraic connectivity of random interdependent networks for certain ranges of inter-layer edge formation probabilities (again, regardless of the intra-layer topologies), showing the importance of the interdependencies between networks in information diffusion dynamics.  Our characterizations were built on a study of the isoperimetric constant of random interdependent and $k$-partite graphs.

There are various interesting avenues for future research, including a deeper investigation of the role of the intra-layer network topologies, and other probability distributions over the inter-layer edges (outside of Bernoulli interconnections).


%

\appendices  
\section{Proof of Lemma \ref{lem:iso}} \label{appendixA}
\begin{IEEEproof}
Suppose the edge formation probability is $p=\frac{\ln n+(r-1)\ln \ln n+x}{(k-1)n}$ where $x=o(\ln \ln n)$ and $x \to \infty$ when $n \to \infty$. We have to show that for any set of vertices of size $m$, $1 \leq m \leq nk/2$, there are at least $m(r-1)+1$ edges that leave the set a.a.s. 

Consider a set $S \subset V_1 \cup V_2 \cup \cdots \cup V_{k}$ with $|S|=m$. Assume that the set $S$ contains $s_i$ nodes from $V_i$ for $1 \leq i \leq k$ (i.e., $|S\cap V_i|=s_i \geq 0$). Define $E_S$ as the event that $m(r-1)$ or fewer edges leave $S$. Note that $|\partial S|$ is a binomial random variable with parameters $\sum_{l=1}^{k} s_l \left( \sum_{t=1, t\neq l}^{k} (n-s_t) \right)$ and $p$.  

We have that
\begin{align} \nonumber
\sum_{l=1}^{k} s_l \left( \sum_{t=1, t\neq l}^{k} (n-s_t) \right)&= \sum_{l=1}^{k} s_l (n(k-1)-m+s_l)\\  \label{lem:iso1}
&= n(k-1)m-m^2 +\sum_{l=1}^{k} s_l^2. 
\end{align}
Then we have 
\begin{align}\label{lem:iso2}
\mathbf{Pr}(E_S)=&\sum_{i=0}^{m(r-1)} {n(k-1)m-m^2 +\sum_{l=1}^{k} s_l^2 \choose i} p^i (1-p)^{n(k-1)m-m^2 +\sum_{l=1}^{k} s_l^2-i}\\ \nonumber
&\leq \sum_{i=0}^{m(r-1)} {n(k-1)m \choose i}  p^i (1-p)^{n(k-1)m-m^2 +\sum_{l=1}^{k} s_l^2-i}\\ \nonumber
&\leq \sum_{i=0}^{m(r-1)} {n(k-1)m\choose i} p^i (1-p)^{n(k-1)m - \frac{(k-1)m^2}{k}-i}.
\end{align}
The first and the second inequality follow from the inequalities $0 \leq m^2 -\sum_{l=1}^{k} s_l^2 \leq \frac{(k-1)m^2}{k}$ which is a straightforward result of the Cauchy-Schwartz inequality in \eqref{cauchysw}. 
Next note that $k \geq 2$ and for $1 \leq i \leq m(r-1)$ and sufficiently large $n$, we have
\begin{align}\nonumber
&\frac{{n(k-1)m \choose i} p^i (1-p)^{n(k-1)m-\frac{(k-1)m^2}{k}-i}}{{n(k-1)m \choose i-1} p^{i-1} (1-p)^{n(k-1)m-\frac{(k-1)m^2}{k}-(i-1)}}\\ \nonumber
&~~~~~~~~~~~~~~~~~=\frac{n(k-1)m-i+1}{i} \times \frac{p}{1-p} \\\nonumber
&~~~~~~~~~~~~~~~~~\geq \frac{n(k-1)m-m(r-1)+1}{m(r-1)} \times \frac{p}{1-p}\\\nonumber
&~~~~~~~~~~~~~~~~~\geq \frac{n(k-1)-r+1}{r-1} \times \frac{p}{1-p}\\\nonumber
&~~~~~~~~~~~~~~~~~> 1. \nonumber
\end{align}
Thus there must exist some constant $R >0$ such that  
\begin{align}\label{lem:iso 3}
\mathbf{Pr}(E_S) &\leq \sum_{i=0}^{ m(r-1)}{n(k-1)m \choose i} p^i (1-p)^{n(k-1)m-\frac{(k-1)m^2}{k}-i}\\\nonumber
&\leq R {n(k-1)m \choose m(r-1)} p^{m(r-1)}(1-p)^{n(k-1)m-\frac{(k-1)m^2}{k}-m(r-1)}.  \nonumber
\end{align}
Define $P_m$ as the probability that there exists a set of nodes $T$ such that $|T|=m$ and $|\partial T|\le m(r-1)$. Then using the inequality ${n \choose m} \le (\frac{ne}{m})^m$ yields
\begin{align} \nonumber
P_m &\leq \sum_{\substack{
|S|=m,\\
 S\subset \cup_{i=1}^{k} V_i
}} \mathbf{Pr}(E_S) \\ \nonumber
&\leq  R {nk \choose m} {n(k-1)m \choose m(r-1)} p^{m(r-1)}  (1-p)^{n(k-1)m-\frac{(k-1)m^2}{k}-m(r-1)}\\  \label{lem:iso 4}
&\leq R \left( \frac{nke}{m} \right)^m \left(\frac{n(k-1)mep}{m(r-1)}\right)^{m(r-1)}  (1-p)^{n(k-1)m - \frac{(k-1)m^2}{k}-m(r-1)}\\ \nonumber
&= R \bigg(\frac{ke^{r}}{(r-1)^{r-1}(1-p)^{r-1}}\frac{n (1-p)^{n(k-1)}(n(k-1)p)^{r-1}}{m(1-p)^{ \frac{(k-1)m}{k}}} \bigg)^m\\ \nonumber
&\le R \left(\frac{c_1 n (1-p)^{n(k-1)}(n(k-1)p)^{r-1}}{m(1-p)^{ \frac{(k-1)m}{k}}} \right)^m. \nonumber
\end{align}
In the last step of the above inequality, $c_1$ is some constant satisfying  
$$
\frac{ke^{r}}{(r-1)^{r-1}(1-p)^{r-1}}\le c_1<\frac{2ke^{r}}{(r-1)^{r-1}},
$$
for sufficiently large $n$.  
Recalling the function $p(n)=\frac{\ln n +(r-1)\ln \ln n+x}{(k-1)n}$ and using the inequality $1-p \leq e^{-p}$ yields
\begin{align*}
P_m &\leq R \bigg(\frac{c_1 n e^{-n(k-1)p}(n(k-1)p)^{r-1}}{m(1-p)^{ \frac{(k-1)m}{k}}} \bigg)^m\\ \nonumber
&= R \bigg( c_1\left(\frac{\ln n +(r-1) \ln \ln n+x}{\ln n}\right)^{r-1} \frac{e^{-x}}{m(1-p)^{\frac{(k-1)m}{k}}} \bigg)^m\\ \nonumber
&\leq R \left(\frac{c_2 e^{-x}}{m(1-p)^\frac{(k-1)m}{k}}\right)^m.
\end{align*}
Due to the fact that $\frac{\ln n +(r-1) \ln \ln n+x}{\ln n} < 2$ for sufficiently large $n$, we can consider a constant upper bound $c_2$ for $c_1\left(\frac{\ln n +(r-1) \ln \ln n+x}{\ln n}\right)^{r-1}$ such that $0<c_2<c_1 2^{r-1}$. Next, we substitute the Taylor series expansion  $\ln (1-p)=-\sum_{i=1}^{\infty} \frac{p^i}{i}$ for $p \in [0,1)$ in the above inequality to obtain
\begin{align*}
P_m &\leq R \left(\frac{c_2 e^{-x} e^{-\frac{(k-1)m}{k} \ln(1-p)}}{m}\right)^m \\
&=R \left(\frac{c_2 e^{-x} e^{\frac{(k-1)mp}{k}} \exp \{\frac{(k-1)m}{k}p^2 \sum_{i=2}^{\infty} \frac{p^{i-2}}{i}\}}{m}\right)^m.
\end{align*}
Since we have $\sum_{i=2}^{\infty} \frac{p^{i-2}}{i} < \sum_{i=2}^{\infty} p^{i-2}=\frac{1}{1-p}$ and $\frac{(k-1)m}{k}p^2 \to 0$ as $n \to \infty$, there must exist a constant $c_3$ such that $0<\frac{(k-1)m}{k}p^2 \sum_{i=2}^{\infty} \frac{p^{i-2}}{i}<c_3<1$ for sufficiently large $n$. Therefore,
\begin{align*}
P_m &\leq R \left(c_2 e^{c_3} \frac{e^{-x} e^{\frac{(k-1)mp}{k}}}{m}\right)^m=R \left(c_4 \frac{e^{-x} e^{\frac{(k-1)mp}{k}}}{m}\right)^m,
\end{align*}
where $0<c_4=c_2 e^{c_3} <\frac{ke^{r+1}2^r}{(r-1)^{r-1}} $. 

Consider the function $f(m)=\frac{e^{\frac{(k-1)mp}{k}}}{m}$. Since $\frac{df}{dm}=\frac{e^{\frac{(k-1)mp}{k}}(\frac{(k-1)mp}{k}-1)}{m^2}$, we have that $\frac{df}{dm}<0$ for $m<\frac{k}{(k-1)p}$ and $\frac{df}{dm}>0$ for $m>\frac{k}{(k-1)p}$. Therefore, $f(m) \leq \max \{f(1), f(nk/2)\}$ for $m \in \{1,2,\ldots, \lfloor nk/2 \rfloor \}$. We have 
\begin{align*}
f(nk/2)=\frac{2e^{\frac{(k-1)nkp}{2k}}}{nk}=\frac{2}{k} e^{(- \ln n+(r-1)\ln \ln n+x)/2}.
\end{align*}
Since $\ln \ln n=o(\ln n)$, we have that $f(nk/2)=o(1)$. Moreover, $1<f(1)=e^{\frac{(k-1)p}{k}}<e$ and thus for sufficiently large $n$ we must have $f(m) \leq f(1)<e$. Therefore,
$$
P_m \leq R (c_4 e^{1-x})^m.
$$
Let $P_0$ be the probability that there exists a set $S$ with size $nk/2$ or less that it is not $r$-reachable. Then by the union bound we have
$$
P_0 \leq \sum_{m=1}^{\lfloor nk/2 \rfloor} P_m \leq \sum_{m=1}^{\infty} R (c_4 e^{1-x})^m=\frac{Rc_4 e^{1-x}}{1-c_4 e^{1-x}}=o(1),
$$
since $x \to \infty$ as $n \to \infty$. Thus we have the required result.
\end{IEEEproof}

\section{Proof of Lemma \ref{lem:iso_bounds}} \label{appendixB}
\begin{IEEEproof}
The inequality $i(G_p) \leq d_{min}$ is clear from the definition of the $i(G_p)$.  We will show that $d_{max} \leq n(k-1)p \left( 1+\sqrt{3} (\frac{\ln n}{(k-1)n p})^{\frac{1}{2}-\epsilon}\right)$ asymptotically almost surely. Let $d_j$ denote the degree of vertex $j$, $1 \leq j \leq kn$. From the definition, $d_j$ is a binomial random variable with parameters $n(k-1)$ and $p$ and thus $\mathbb{E}[d_j]=n(k-1)p$. Then, for any $0 < \beta \leq \sqrt3$ we have \cite{Mitzenmacher,TAC}
\begin{equation}
\label{chernof}
\mathbf{Pr}(d_j \geq (1+\beta)\mathbb{E}[d_j]) \leq e^{-\frac{\mathbb{E}[d_j] \beta^2}{3}}.
\end{equation}
Choose $\beta=\sqrt{3} (\frac{\ln n}{(k-1)n p})^{\frac{1}{2}-\epsilon}$, which is at most $\sqrt3$ for $p$ satisfying the conditions in the lemma and for sufficiently large $n$. Substituting into equation \eqref{chernof}, we have
\begin{equation*}
\mathbf{Pr}(d_j \geq (1+\beta)\mathbb{E}[d_j]) \leq e^{-(\ln n)(\frac{\ln n}{(k-1)n p})^{-2\epsilon}}.
\end{equation*}

The probability that $d_{max}$ is higher than $(1+\beta)\mathbb{E}[d_j]$ equals the probability that at least one of the vertices has degree higher than  $(1+\beta)\mathbb{E}[d_j]$, which by the union bound is upper bounded by
\begin{align*}
\mathbf{Pr}(d_{max} \geq (1+\beta)\mathbb{E}[d_j]) &\leq kn \mathbf{Pr}(d_j \geq (1+\beta)\mathbb{E}[d_j])\\
&\leq k e^{(\ln n) -(\ln n)(\frac{\ln n}{(k-1)n p})^{-2\epsilon}}\\
&\leq k e^{(\ln n)(1- (\frac{\ln n}{(k-1)n p})^{-2\epsilon})}.
\end{align*}
Since the right hand-side of the above inequality goes to zero as $n \to \infty$ for $p$ satisfying the condition in the lemma, we conclude that 
$$
d_{max} \leq n(k-1)p \left( 1+\sqrt{3} \left(\frac{\ln n}{(k-1)n p}\right)^{\frac{1}{2}-\epsilon}\right),
$$
asymptotically almost surely.

Next we aim to prove the lower-bound for $i(G_p)$ in \eqref{prop_e_1}. In order to prove this, we show that for any set of vertices of size $m$, $1 \leq m \leq nk/2$, there are at least $\alpha m n p$ edges that leave the set, for some constant $\alpha$ that we will specify later and probability $p$ satisfying $\lim_{n \to \infty} \frac{\ln n}{(k-1)n p} <1$. 

Consider a set $S \subset V_1 \cup V_2 \cup \cdots \cup V_{k}$ with $|S|=m$. 
Assume that the set $S$ contains $s_i$ nodes from $V_i$ for $1 \leq i \leq k$ (i.e., $|S\cap V_i|=s_i \geq 0$). Define $E_S$ as the event that $\alpha m n p$ or fewer edges leave $S$. Note that $|\partial S|$ is a binomial random variable with parameters $\sum_{l=1}^{k} s_l \left( \sum_{t=1, t\neq l}^{k} (n-s_t) \right)$ and $p$.  Similarly to the equality \eqref{lem:iso1} and inequality \eqref{lem:iso2}, we have that 
\begin{align}\label{newES}
\mathbf{Pr}(E_S) &\leq \sum_{i=0}^{\lfloor \alpha m n p \rfloor} {n(k-1)m\choose i} p^i (1-p)^{n(k-1)m - \frac{(k-1)m^2}{k}-i}.
\end{align}
Next note that $k \geq 2$ and for $1 \leq i \leq \lfloor \alpha m n p \rfloor$,
\begin{align}\nonumber
&\frac{{n(k-1)m \choose i} p^i (1-p)^{n(k-1)m-\frac{(k-1)m^2}{k}-i}}{{n(k-1)m \choose i-1} p^{i-1} (1-p)^{n(k-1)m-\frac{(k-1)m^2}{k}-(i-1)}}\\ \nonumber
&~~~~~~~~~~~~~~~~~=\frac{n(k-1)m-i+1}{i} \times \frac{p}{1-p} \\\nonumber
&~~~~~~~~~~~~~~~~~\geq \frac{n(k-1)m-\alpha m n p+1}{\alpha m n p} \times \frac{p}{1-p}\\\nonumber
&~~~~~~~~~~~~~~~~~\geq \frac{k-1-\alpha p}{\alpha} \times \frac{1}{1-p}\\\nonumber
&~~~~~~~~~~~~~~~~~\geq \frac{1}{\alpha}, \nonumber
\end{align}
for $\alpha <1$ which will be satisfied by our eventual choice for $\alpha$.  

Now let $P_m$ denote the probability of the event that there exists a set of size $m$ with $\lfloor \alpha m n p \rfloor$ or fewer number of edges leaving it. Then there must exist some constant $R >0$ such that by the same procedure as in inequalities \eqref{lem:iso 3} and \eqref{lem:iso 4}, we have 
\begin{align}\nonumber
P_m &\leq R \left( \frac{nke}{m} \right) ^{m} \left( \frac{n(k-1)mep}{\alpha m n p} \right) ^{\alpha m n p}  (1-p)^{n(k-1)m- \alpha m n p-\frac{(k-1)m^2}{k}}\\  \nonumber
&\leq R \left( \frac{(k-1)e}{\alpha}\right)^{\alpha m n p} e^{m \ln \left(\frac{nke}{m}\right)}  e^{-p(n(k-1)m -\alpha m n p-\frac{(k-1)m^2}{k})}\\
&= R e^{m h(m)},\label{shs}
\end{align}
where 
\begin{align}\nonumber
h&(m)=\alpha n p+\alpha n p \ln (k-1)-\alpha n p \ln \alpha+ \ln\left(\frac{nke}{m}\right)-p(n(k-1)-\alpha n p-\frac{(k-1)m}{k})\\ 
&=1+\ln k+\frac{(k-1)pm}{k}-\ln m \label{hm}+np\left(\underbrace{\alpha +\alpha \ln (k-1)-\alpha \ln \alpha +\frac{\ln n}{np}-(k-1)+\alpha p}_\text{$\Gamma(\alpha)$}\right).
\end{align}
Since $\frac{\partial h(m)}{\partial m}=\frac{(k-1)p}{k}-\frac{1}{m}$ is negative for $m < \frac{k}{(k-1)p}$ and positive for $m > \frac{k}{(k-1)p}$, we have
\begin{align*}
h(m) &\leq \max \{ h(1), h(nk/2)\}\\
&\leq \max \Big\{1+\ln k+\frac{(k-1)p}{k}+np \Gamma(\alpha),1+\ln 2+np\left(\Gamma(\alpha)+\frac{(k-1)}{2}-\frac{\ln n}{np}\right)\Big\}.
\end{align*}

From \eqref{hm}, $\frac{\partial \Gamma(\alpha)}{\partial \alpha}=\ln (k-1) -\ln \alpha+p>0$ and thus $\Gamma(\alpha)$ is an increasing function in $\alpha$ for $\alpha <(k-1)$, with $\Gamma(0)=\frac{\ln n}{np}-(k-1)$ which is negative and bounded away from 0 for sufficiently large $n$ (by the assumption on $p$ in the lemma). Therefore, there exists some sufficiently small positive constant $\alpha$ such that $h(m)\leq -\alpha n p $ for sufficiently large $n$. Thus \eqref{shs} becomes $P_m \leq R e^{-\alpha m n p}$ for sufficiently large $n$.

The probability that $i(G_p) \leq \alpha n p$ is upper bounded by the sum of the probabilities $P_m$ for $1 \leq m \leq nk/2$. Using the above inequality, we have
\begin{align*}
\mathbf{Pr}(i(G_p) \leq \alpha n p) \leq \sum_{m=1}^{nk/2} P_m &\leq R \sum_{m=1}^{nk/2} e^{-\alpha m n p}\\
& \leq R \sum_{m=1}^{\infty} e^{-\alpha m n p}\\
&= R \frac{e^{-\alpha n p}}{1-e^{-\alpha n p}},
\end{align*}
which goes to 0 as $n \to \infty$. Therefore, we have $i(G_p) \geq \alpha n p$ asymptotically almost surely. 
\end{IEEEproof}

%

\bibliographystyle{IEEEtran}
\bibliography{refs}

\end{document}